\documentclass[12pt]{article}
\usepackage{hyperref}
\usepackage{graphicx}
\usepackage{amsmath}
\usepackage{amssymb}
\usepackage{makeidx}
\usepackage{fullpage}

\begin{document}


\baselineskip 0.7cm

\begin{titlepage}

\begin{flushright}
IPMU 09-0143\\
CALT-68-2757
\end{flushright}

\vskip 1.35cm

\begin{center}


{\large\bf Cascade Events at IceCube+DeepCore as a Definitive Constraint 
on the Dark Matter Interpretation of the PAMELA and Fermi Anomalies}

\vskip 1.2cm
Sourav K. Mandal$^{1,2}$, Matthew R. Buckley$^{3}$, Katherine Freese$^{4}$, 
Douglas Spolyar$^{5,6}$, and Hitoshi Murayama$^{1,2,7}$
\vskip 0.4cm
{\it 
$^{1}$Department of Physics, University of California, Berkeley, CA 94720, USA\\
$^{2}$IPMU, University of Tokyo, 5-1-5 Kashiwa-no-ha, Kashiwa, 
Japan 277-8568\\
$^{3}$Department of Physics, California Institute of Technology, Pasadena, 
CA 91125, USA\\
$^{4}$Michigan Center for Theoretical Physics, Physics Department, University of
Michigan,
Ann Arbor, MI 48109, USA\\
$^{5}$Center for Particle Astrophysics, Fermi National Accelerator Laboratory,\\
Batavia, IL 60510, USA\\
$^{6}$University of California, Santa Cruz, Physics Department, Santa Cruz, CA
95064, USA\\
$^{7}$Theoretical Physics Group, LBNL, Berkeley, CA 94720, USA
}

\end{center}

\vskip 1.5cm

\begin{abstract}
Dark matter decaying or annihilating into $\mu^+\mu^-$ or 
$\tau^+\tau^-$ has been proposed as an explanation for the $e^\pm$ anomalies
reported by PAMELA and Fermi.  Recent analyses show that IceCube, supplemented
by DeepCore, will be able to significantly constrain the parameter space of decays
to $\mu^+\mu^-$, and rule out decays to $\tau^+\tau^-$ 
and annihilations to $\mu^+\mu^-$ in less than five years of running.  
These analyses rely on measuring track-like events in 
 IceCube+DeepCore from down-going $\nu_\mu$. In this paper we show that by 
 instead measuring cascade events, which are induced by all neutrino flavors, 
 IceCube+DeepCore can rule out decays to $\mu^+\mu^-$ in only three years of running, 
 and rule out decays to $\tau^+\tau^-$ and annihilation to $\mu^+\mu^-$ 
 in only one year of running.  These constraints are highly 
 robust to the choice of 
 dark matter halo profile and independent of dark matter-nucleon cross section.
\end{abstract}

\end{titlepage}
\setcounter{page}{2}



\section{Introduction}
The existence of dark matter has been established by numerous observations.
Although it constitutes most of the matter in the universe~\cite{WMAP5}, 
the nature of
dark matter remains largely unknown.  One widely held possibility is that it is
a new fundamental particle produced in the early universe and present today 
as a thermal relic.  Among the best motivated of these are the so-called
weakly interacting massive particles, or ``WIMPs''
(for reviews, see Refs.~\cite{review:jungman,review:bertone}) which are
predicted to be undergoing annihilations~\cite{ann:sun,ann:earth,ann:halo,ann:dsph}
and possibly decays~\cite{wimp:decay} in the current epoch.

Recently, the
instruments PAMELA~\cite{Adriani:2008zr}, ATIC~\cite{ATIC}, 
PPB-BETS~\cite{Torii:2008xu}, and Fermi~\cite{Abdo:2009zk} have observed features in the 
cosmic-ray $e^\pm$ spectrum and a positron fraction that are inconsistent with 
known backgrounds.  While these anomalies may be due to unidentified 
astrophysical sources~\cite{pulsars}, one exciting possibility is that they are 
due to the 
decay~\cite{Chen:2008yi,Chen:2008dh,Yin:2008bs,
  Ishiwata:2008cv,Shirai:2009kh,Ibarra:2009bm,Ibe:2009en} or 
  annihilation~\cite{Barger:pmodel,Cholis:pmodel,Bergstrom:pmodel,
  Sommerfeld:cirelli,
  Various:pmodels,Sommerfeld:nima}
   of dark matter particles into standard model states.  Even in the case that
anomalies are not caused by new physics in the dark sector, the constraints
are generally applicable to dark matter models.

In order for dark matter to explain the anomalies, the products of
decay or annihilation must be 
primarily leptonic.  In either case, the Fermi observation
gives the most precise preferred region in mass and lifetime/cross-section; 
for decays, only $\mu^+\mu^-$ and $\tau^+\tau^-$ final states 
fit the data, while for annihilations a small region of 
$e^+e^-$ is also permitted~\cite{Meade:fermi}.
For the allowed parameter space, decays or annihilations into hadrons and
weak and Higgs bosons are severely limited.  Decays are constrained by the 
Fermi observation of the isotropic diffuse gamma-ray flux~\cite{Chen:fermiiso}, 
and annihilations are
constrained by the production of antiprotons~\cite{Adriani:2008zq}.

Explaining the anomalies with annihilations has an additional challenge.
In order to match the observed rates, the cross-section required is $10^3$--$10^4$ times that 
expected for thermal production of the dark matter in the early universe.
  This necessitates nonthermal production mechanisms or low-velocity enhancements
to the cross-section, such as the
Sommerfeld enhancement~\cite{Sommerfeld:cirelli,Sommerfeld:nima}
 or the Breit-Wigner enhancement~\cite{Feldman:breit,Ibe:breit}.
Moreover, high cross section annihilations to leptonic states are 
tightly constrained by synchrotron radio emissions from the galactic center, 
although this constraint can be evaded if the true galactic dark matter halo
profile is much less steep at the galactic center than 
benchmark profiles~\cite{Bertone:gcradio,Meade:fermi}.

For either decays or annihilations, neutrino observations may provide the 
strongest constraints --- or the most promising corroborating signatures ---
 for dark matter to be the source of the anomalies.  
 With the exception of the disfavored $e^+e^-$ channel,
 the required leptonic final states decay into neutrinos.
 These travel undeflected
from their sources, eliminating any uncertainties in modeling their propagation.  
Moreover,
astrophysical sources that may explain the anomalies are not expected
to produce a large flux of neutrinos.  Pulsars, for example, would generate at most 
${\cal O}(1)$ events/year at a full km-scale detector, and only at
energies greater than 10 TeV~\cite{pulsarbkg}.

There are, however, challenges to constraining dark matter models with neutrinos.  
As they are observed only by collecting Cherenkov light from induced particle showers 
or from secondary muons, angular resolution is poor compared to gamma-ray
observatories.  Also, there are 
significant backgrounds from atmospheric muon and 
 neutrino fluxes.  However, if these backgrounds can be controlled, the poor
 angular resolution need not be a barrier; indeed by integrating the flux
 over a large area of the sky, the resulting constraint is much less 
 sensitive to the choice of dark matter 
 profile~\cite{beacom:anncosm,yuksel:anngal}.  
 Moreover, by observing the 
 galactic and extragalactic diffuse dark matter, rather than any that may 
 have been captured by the Sun or the Earth, any constraints will be independent of
 the dark matter-nucleon cross sections, which are related to final states in
 a model-specific way~\cite{dmcapture}.
 
Recent analyses show the power of neutrino constraints, using various strategies
to reduce the effect of atmospheric backgrounds.  The Super-Kamiokande 
observatory resides in the northern hemisphere, facing away from the galactic
center, minimizing atmospheric backgrounds.  Measurements of upward-going
muons place a limit on the flux of galactic $\nu_\mu$, providing a robust 
constraint that eliminates annihilations to $\tau^+\tau^-$ as a source of the $e^\pm$
anomalies~\cite{superk,Meade:fermi}.  
The IceCube observatory, on the other hand, resides
at the South Pole where down-going atmospheric fluxes are coincident with the 
neutrinos from the galactic center.  The overwhelming background
of atmospheric muons can be suppressed by event selection to establish
an isotropic diffuse flux 
limit~\cite{kowalski:diff,gerhardt:amanda,kiryluk:icecube}, 
but this limit only starts at a high energy 
threshold ${\cal O}(10\;{\rm TeV})$, and yields a relatively weak
constraint as we will show.

Currently under construction is DeepCore~\cite{deepcore}, 
an in-fill of the IceCube detector
which will use the outer instrumented volume as a veto on downward-going muons
to a level of one part in $10^6$~\cite{deepcorestatus}.  
This will allow
the galactic neutrino flux to be measured and compared against the 
atmospheric neutrino flux, providing a constraint on dark matter decays 
and annihilations.  
Recent work~\cite{dc:ann,dc:decay} by some of the authors
 shows that IceCube+DeepCore 
will be able to significantly constrain the parameter space of decays
to $\mu^+\mu^-$, and rule out decays to $\tau^+\tau^-$ 
and annihilation to $\mu^+\mu^-$ as possible sources of the anomalies
 in less than five years of running.

\begin{figure}[t!]
\begin{center}
\includegraphics[width=0.49\linewidth]{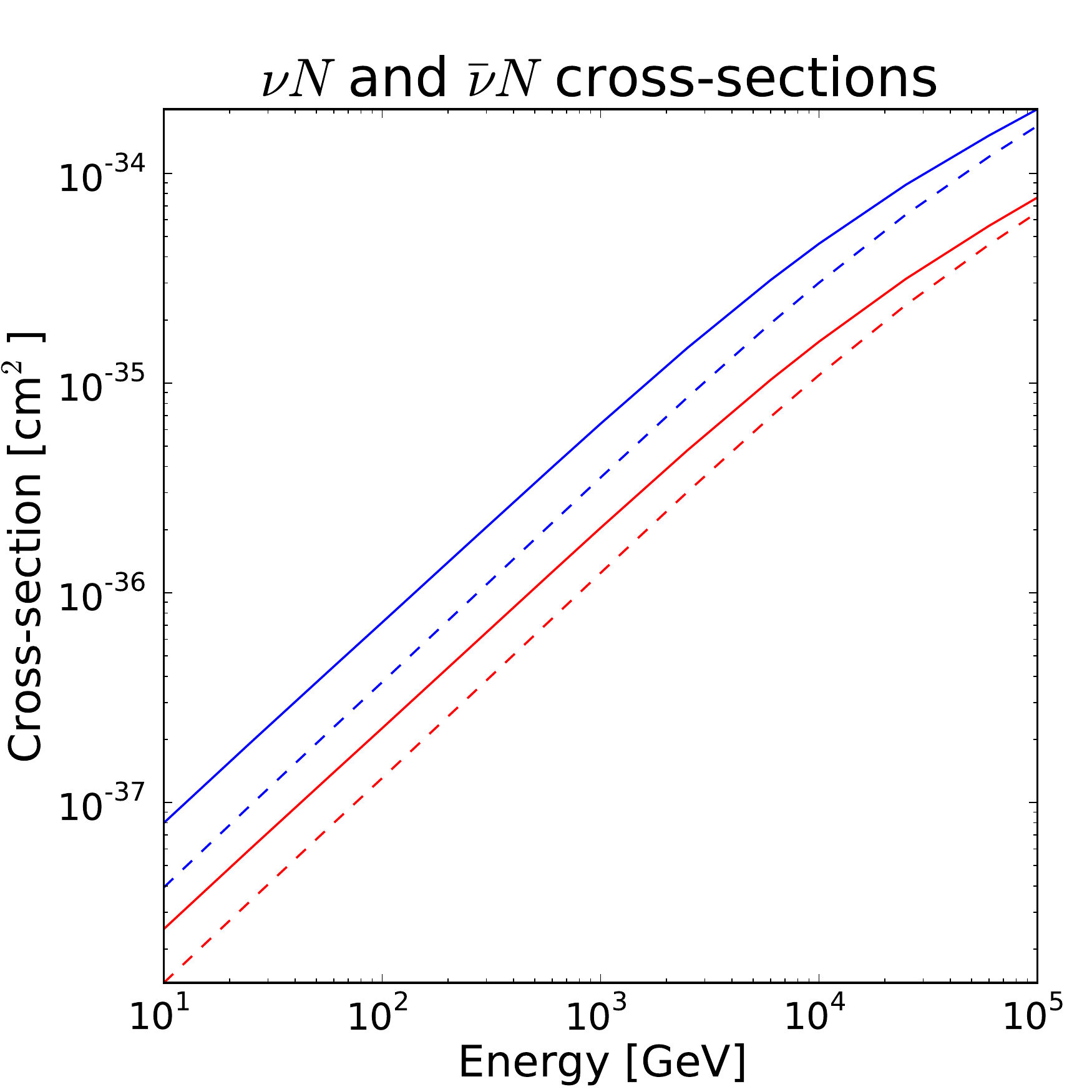}
\end{center}
\caption{Cross sections for (anti)neutrino-nucleon interactions,
given by Ref.~\cite{nuNcs}.  The blue lines
are for charged-current interactions, and the red lines are for 
neutral-current interactions.  Solid lines are for neutrinos and dashed lines
are for antineutrinos.}
\label{nuNcs}
\end{figure}

Recently, IceCube+DeepCore has demonstrated in 
simulations the ability to distinguish between track-like events, which are due 
 to the charged-current interactions of
$\nu_\mu$, and cascade events, which are induced by $\nu_{e,\tau}$ through
charged-current interactions and by all neutrino flavors through 
neutral-current interactions~\cite{middell:casc,grant:nutau}.
 This is very useful for constraining dark matter neutrino fluxes because
 $\nu_\mu$ is the dominant flavor of atmospheric neutrinos above 
 40~GeV~\cite{numunuebkg,nutaubkg}.
 The neutrino-nucleon cross sections are the same for all flavors, 
 so signal would be considerably enhanced, while background 
 would be reduced because $\nu_\mu$
 only creates cascade events through the neutral-current interaction, which 
 is lower in cross section than the charged-current 
 interaction~\cite{nuNcs,beacom:casc,beacom:anncosm}
 (see Fig.~\ref{nuNcs}).  Moreover, cascade events are easy to distinguish
 from the tracks caused by any atmospheric muons that are not vetoed by the
 outer volume.

 In this paper we show that observation of cascade events at IceCube+DeepCore
 can enhance the neutrino constraints on dark matter, and rule out
 (or corroborate) dark matter decays to 
 $\mu^+\mu^-$ or $\tau^+\tau^-$ and 
 annihilation to $\mu^+\mu^-$ as sources of the observed $e^\pm$ cosmic-ray
 anomalies in a much shorter time compared to searches that rely solely on
 track-like events.
 
\section{Analysis}
For brevity, in the expressions below we write only the terms for neutrinos
and not the terms for antineutrinos.
The terms are identical, except 
replacing ``$\nu$'' by ``$\bar\nu$''.
\subsection{Galactic flux signal and background}
The flux of neutrinos from the galactic dark matter halo is given by 
\begin{equation}
\frac{d\Phi_{\nu_i}}{dEd\Omega}
=\frac{1}{4\pi}(r_\odot\rho_\odot)\frac{1}
{m \tau}
\frac{dN_{\nu_i}}{dE}
J_1(\Delta\Omega)
\end{equation}
for decays and by
\begin{equation}
\frac{d\Phi_{\nu_i}}{dEd\Omega}
=\frac{1}{8\pi}(r_\odot\rho_\odot^2)\frac{1}
{m^2}\left<\sigma v\right>
\frac{dN_{\nu_i}}{dE}
J_2(\Delta\Omega)
\end{equation}
for annihilations, where $r_\odot=8.5$ kpc is the distance from the Sun to the
galactic center~\cite{rsun}, 
$\rho_\odot=0.3\;{\rm GeV}\,{\rm cm}^{-3}$ is the dark matter
density in the solar neighborhood, $m$ is the dark matter mass,
and $\tau$ and $\left<\sigma v\right>$ are the dark 
matter lifetime and thermally averaged cross section respectively. 
$J_n$ is the line-of-sight (los) integral through the halo profile:
\begin{equation}
J_n(\Delta\Omega)=\frac{1}{\Delta\Omega}\int_{\Delta\Omega} d\Omega
\int_{\rm l.o.s.}\frac{ds}{r_\odot}\left(\frac{\rho}{\rho_\odot}\right)^n
\end{equation}
where $\Delta\Omega$ is the region of sky observed. 
In this analysis we use the Navarro-Frenk-White (NFW) halo
profile~\cite{Navarro:1996gj}
\begin{equation}
\rho(r)=\rho_\odot\left(\frac{r_\odot}{r}\right)
\left(\frac{1+r_\odot/r_s}{1+r/r_s}\right)^2
\end{equation}
with $r_s=20$ kpc. 

The neutrino source spectra $dN_{\nu_i}/dE$ for 
flavors $i$ are
given by {\tt PYTHIA}~\cite{Sjostrand:2006za} simulation.
Assuming
 $\sin^2 2\theta_{12}:\sin^2 2\theta_{13}:\sin^2 2\theta_{23}\approx 1:1:0$, 
 the flavor distribution will be $1:1:1$ as the neutrinos reach the Earth,
 having traveled a variety of long distances across the galaxy and being well 
 mixed
 through vacuum oscillations.

As discussed in the Introduction, the only background to
cascade events is atmospheric neutrinos after the
veto suppresses the background of atmospheric muons to one part in $10^{6}$
and event selection is used to eliminate the rest.  The
$\nu_\mu$ and $\nu_e$ fluxes are given by Ref. \cite{numunuebkg}, where 
the $\nu_\mu$ flux agrees well with AMANDA observation~\cite{amanda:numu}.
At low zenith angles, the flux of background $\nu_\mu$ is $\sim 20$ times greater than 
the flux of $\nu_e$ from 40 GeV to 100 TeV; it is $\sim 1000$ times greater
than the flux of $\nu_\tau$ (see Ref.~\cite{nutaubkg}).  
However, below 40 GeV, especially at high zenith
angles, the atmospheric fluxes of the three flavors only differ by ${\cal O}(1)$
due to flavor mixing.
Since modeling the background (and signal) below 40 GeV would require simulating flavor
mixing as the neutrinos propagate through the limb of the Earth, for this 
analysis we simply set an energy cutoff of $E_{\rm thresh}=40$ GeV.

To obtain the event rates due to the galactic and background fluxes, 
we first set energy bins for dark matter of mass $m$ to be
$[{\rm max}(E_{\rm thresh}, m/10), m/2]$ for 
decays and $[{\rm max}(E_{\rm thresh}, m/5), m]$ 
for annihilations.  Note that the bin width is much greater than the energy 
resolution, $\log_{10}(E_{\rm max}/E_{\rm min})\simeq 0.4$ for track-like events
and $\log_{10}(E_{\rm max}/E_{\rm min})\simeq 0.18$ 
for cascade events~\cite{deepcorestatus}.
We then integrate the flux times the effective area over the energy for 
each bin.

For track-like events due to $\nu_\mu$, the event rate is
\begin{equation}
\label{eq:track}
\Gamma_{\rm tr.}=\int d\Omega\int_{E_{\rm min}}^{E_{\rm max}}dE\;
 \rho_{\rm ice}N_A V_{\rm tr.}\left(
[\sigma_{\nu N}(E)]_{\rm CC}\,\frac{d\Phi_{\nu_\mu}}{dEd\Omega}
\right)
\end{equation}
where $\rho_{\rm ice}=0.9\;{\rm g}\,{\rm cm}^{-3}$ is the density of ice, 
$N_A=6.022\times 10^{23}\;{\rm g}^{-1}$ is Avogadro's number (to convert
mass to number of nucleons), $[\sigma_{\nu N}(E)]_{\rm CC}$ 
is the neutrino-nucleon cross section for the charged-current interaction,
and $V_{\rm tr.}\approx 0.04\;{\rm km}^3$ is the effective volume of
the detector for track-like events~\cite{deepcorestatus}. 
Note that we do not add the residual atmospheric muon background to the
background of track-like events due to the uncertainty in its value.

For cascade events
we use the instrumented volume 
$V_{\rm casc}\approx 0.02\;{\rm km}^3$~\cite{deepcorestatus,mcdonald:volume},
the charged-current interaction for $\nu_{e,\tau}$, and the
neutral-current interaction for all flavors to obtain
\begin{equation}
\label{eq:casc}
\Gamma_{\rm casc.}=\int d\Omega\int_{E_{\rm min}}^{E_{\rm max}}dE\;
 \rho_{\rm ice}N_A V_{\rm casc.}
 \left([\sigma_{\nu N}(E)]_{\rm CC}\,\frac{d\Phi_{\nu_{e,\tau}}}{dEd\Omega}
 +[\sigma_{\nu N}(E)]_{\rm NC}\,\frac{d\Phi_{\nu_{e,\mu,\tau}}}{dEd\Omega}\right)\;.
\end{equation}
Unlike track-like events, cascade events are well contained,
so the effective volume for their detection will vary little with energy.
Also, we assume that in neutral-current interactions all the energy of 
the neutrino goes into the cascade.  Taking partial energy transfer into
account would yield a 
modest improvement in significance, since most of the signal is from 
$\nu_{e,\tau}$ charged-current interactions but most of the background is from
$\nu_\mu$ neutral-current interactions.\footnote{Inelasticity curves are given
in Ref.~\cite{nuNcsold}, but only for energies 10~TeV and higher.}

Finally, because 
the pointing capability for cascades is approximately
$50^\circ$~\cite{deepcorestatus,middell:casc} and the pointing capability
for track-like events has yet to be established, 
we integrate over the $2\pi$ field of view around the 
galactic center.  
As mentioned before, this should provide a constraint
which is robust to changes in dark matter halo profile.  Specifically, 
the fractional change between the NFW profile used here and the 
much less steep
isothermal profile for $J_1(2\pi)$ is ${\cal O}(10^{-3})$, and 
for $J_2(2\pi)$ is ${\cal O}(0.1)$.

\subsection{Extragalactic isotropic diffuse flux}
 For comparison to the DeepCore constraints from down-going fluxes,
 we calculate the constraint from the isotropic diffuse flux limit for AMANDA-II
 from track-like events~\cite{gerhardt:amanda} and the 
 projected limit for IceCube from cascade events~\cite{kiryluk:icecube}.
 
The main contribution from dark matter decay
to the isotropic diffuse flux is that from extragalactic dark matter 
residing on cosmological scales.  The flux from 
the decay of cosmological dark matter is given by the
 previously derived~\cite{Ibarra:exgal,Chen:fermiiso} formula
\begin{equation}
\left(\frac{d\Phi_{\nu_i}}{dEd\Omega}\right)^{\rm (ex.)}
=\frac{c}{4\pi}
\frac{\Omega_{\rm DM}\rho_c}{H_0\Omega_{\rm M}^{1/2}}\frac{1}{m \tau}
\int_1^\infty dy 
\frac{y^{-3/2}}{\sqrt{1+\Omega_\Lambda/\Omega_{\rm M}y^{-3}}}
\left(\frac{dN_{\nu_i}}{d(Ey)}\right)
\end{equation}
where $y\equiv 1+z$, with $z$ being the redshift, 
$H_0=71.9\;{\rm
  km}\,{\rm s}^{-1}\,{\rm Mpc}^{-1}$ is the Hubble constant,
$\rho_c=3H_0^2/(8\pi G_N)$ is the critical density, $\Omega_{\rm
  M}=0.258$, $\Omega_{\rm DM}=0.214$, and
$\Omega_\Lambda=0.721$ are, respectively, the total matter, dark matter, and dark
energy densities divided by the critical density~\cite{WMAP5}. 
The isotropic diffuse flux from the annihilations of cosmological
dark matter is too small to be relevant,
since the density of dark matter on cosmological scales is very low, and
the flux is suppressed by another power of $\rho_c/m$.

Because of the loss of signal due to event selection and the 
presence of background
fluxes, both the AMANDA-II limit and the projected IceCube limit are only valid
at energies greater than
$\sim20$ TeV.  
Below these thresholds we add the atmospheric background flux to these limits,
and use these total fluxes to calculate the constraints on the dark matter 
lifetime.

\clearpage

\section{Results}
\begin{figure}[t!]
\begin{center}
\includegraphics[width=0.49\linewidth]{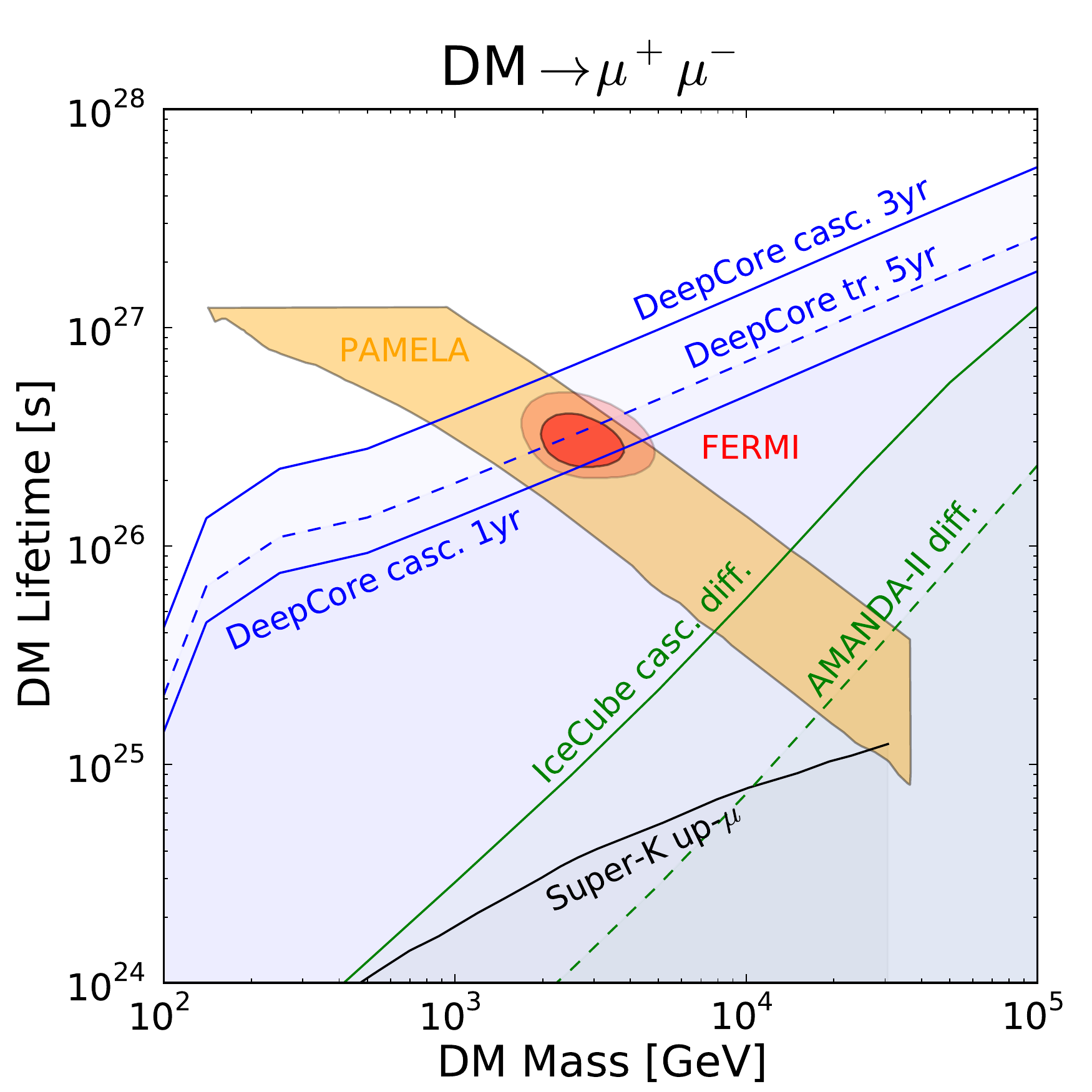}
\includegraphics[width=0.49\linewidth]{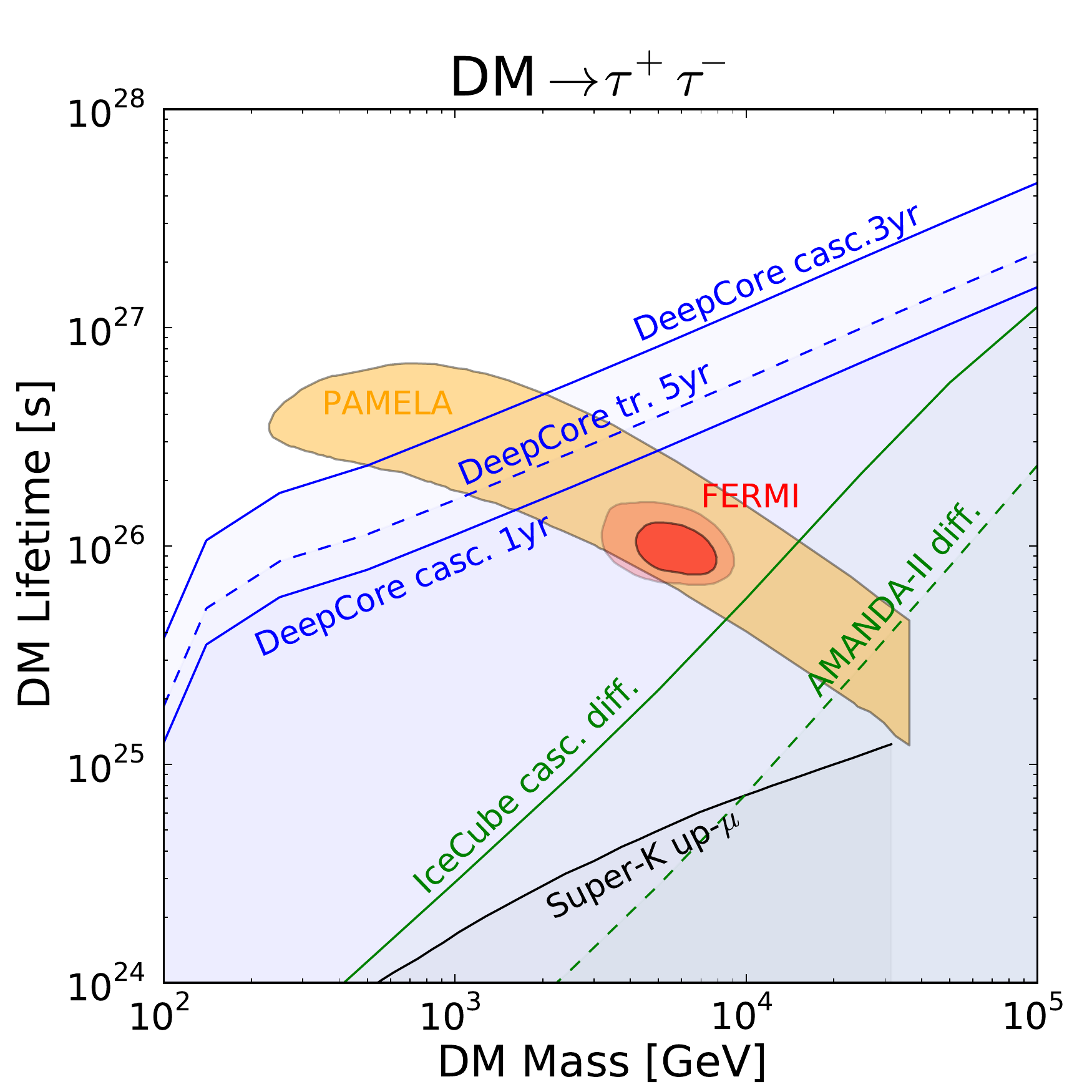}
\end{center}
\caption{Constraints for decay to $\mu^+\mu^-$ (left) and $\tau^+\tau^-$
(right); the regions below the contours are excluded.
The black contour (``Super-K up-$\mu$'') is the 
Super-Kamiokande limit to $3\sigma$ from
up-going muons, the orange band
is the PAMELA-preferred region, and the red ellipses 
are the Fermi-preferred region; these three are given 
by Ref.~\cite{Meade:fermi}.  The dashed green
line (``AMANDA-II diff.'') 
is the constraint to 90\% confidence from the AMANDA-II limit on the isotropic diffuse flux
of $\nu_\mu$, and the solid green line (``IceCube casc. diff.'') 
is the constraint to 90\% confidence from the projected IceCube
limit on the isotropic diffuse flux using cascade events.
The dashed blue line (``DeepCore tr. 5yr'') is the constraint to $2\sigma$
from IceCube+DeepCore for $\nu_\mu$ track-like 
events after five years of running, and the solid blue lines 
are the constraints to $2\sigma$
for all-flavor cascade events after one year (``DeepCore casc. 1yr'') 
and three years (``DeepCore casc. 3yr'') 
of running.}
\label{dec}
\end{figure}
The results for dark matter decays are shown in Fig.~\ref{dec}; the regions
below the contours are excluded.
The black contour (``Super-K up-$\mu$'') is the 
Super-Kamiokande limit to $3\sigma$ from
up-going muons discussed in the Introduction.
The orange band
is the preferred region to fit the PAMELA $e^\pm$ anomaly, and the red ellipses 
are the preferred region 
to fit the Fermi $e^\pm$ anomaly.  
These three regions are given by Ref.~\cite{Meade:fermi} up to mass 30~TeV 
and lifetime $10^{27}$~s.

The dashed green
line (``AMANDA-II diff.'') 
is the constraint to 90\% confidence
from the AMANDA-II limit on the isotropic diffuse flux
of $\nu_\mu$, and the solid green line (``IceCube casc. diff.'') 
is the constraint to 90\% confidence from the projected IceCube
limit on the isotropic diffuse flux using cascade events.  Since the flux 
from cosmological dark matter is weak to begin with, contributing only $1\%$
of the total flux over the $2\pi$ facing the galactic center, we see that 
it is 
quickly overwhelmed by the atmospheric flux below $\sim 40$ TeV.  Nonetheless,
due to the lower background of atmospheric $\nu_e$ compared to $\nu_\mu$
at these high energies, using cascade events improves the constraint by 
a factor of $\sim 5$.

The dashed blue line (``DeepCore tr. 5yr'') is the constraint to $2\sigma$
from IceCube+DeepCore for track-like 
events after five years of running, and the solid blue lines 
are the constraints to $2\sigma$
for cascade events after one year (``DeepCore casc. 1yr'') 
and three years (``DeepCore casc. 3yr'') 
of running.  We see that for the $\mu^+\mu^-$
final state, while track-like events can only reduce the available Fermi-preferred
parameter space in five years, cascade events can rule it out altogether in only three years.
Similarly for the $\tau^+\tau^-$ final state, track-like events can rule out 
the parameter space in less than five years, but with cascade events it will
only take one year.  Note the weakening of the constraints below $m=250$ GeV, where the 
energy per final-state particle is less than 125 GeV.  This is caused by the 
energy cutoff at 40 GeV.

\begin{figure}[t!]
\begin{center}
\includegraphics[width=0.49\linewidth]{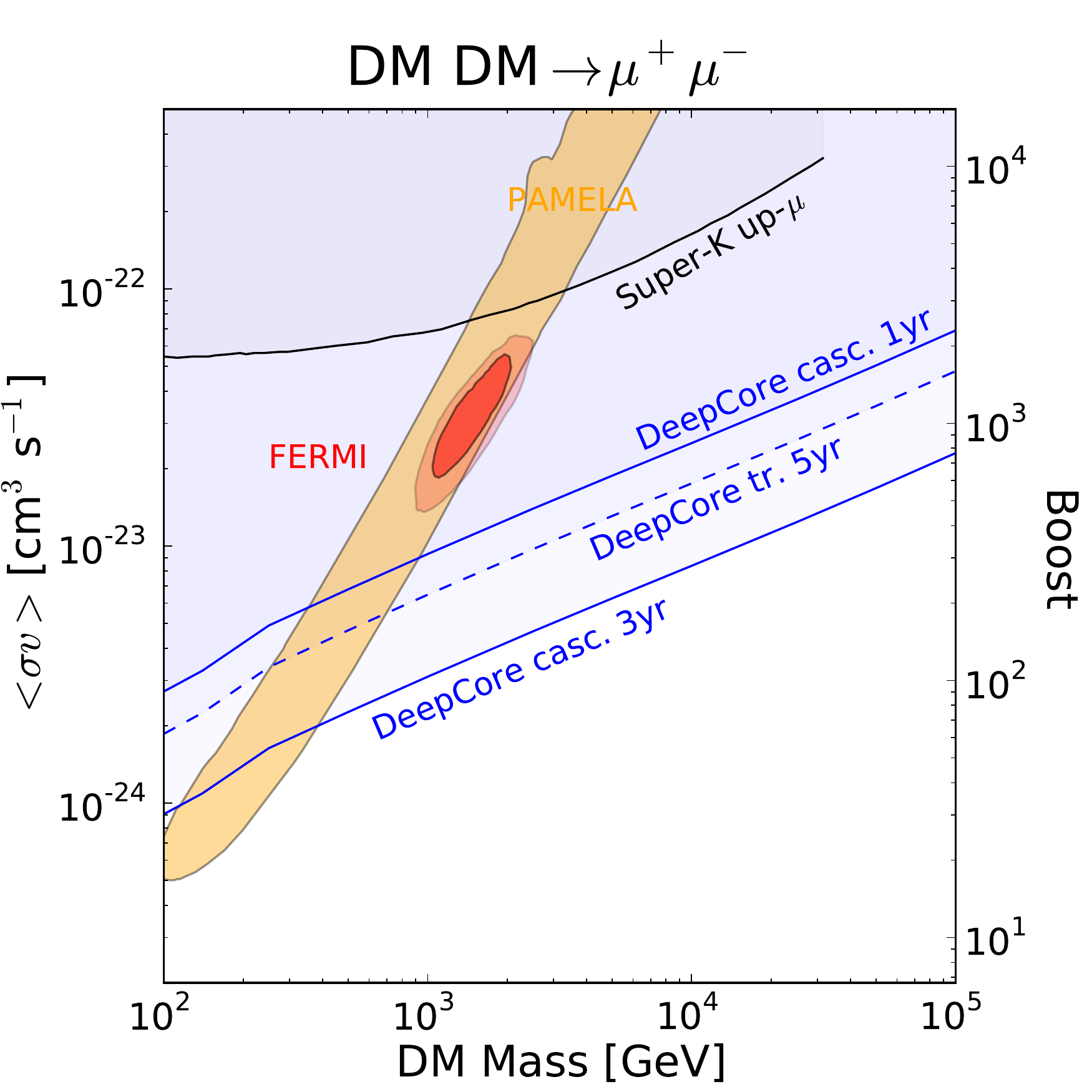}
\includegraphics[width=0.49\linewidth]{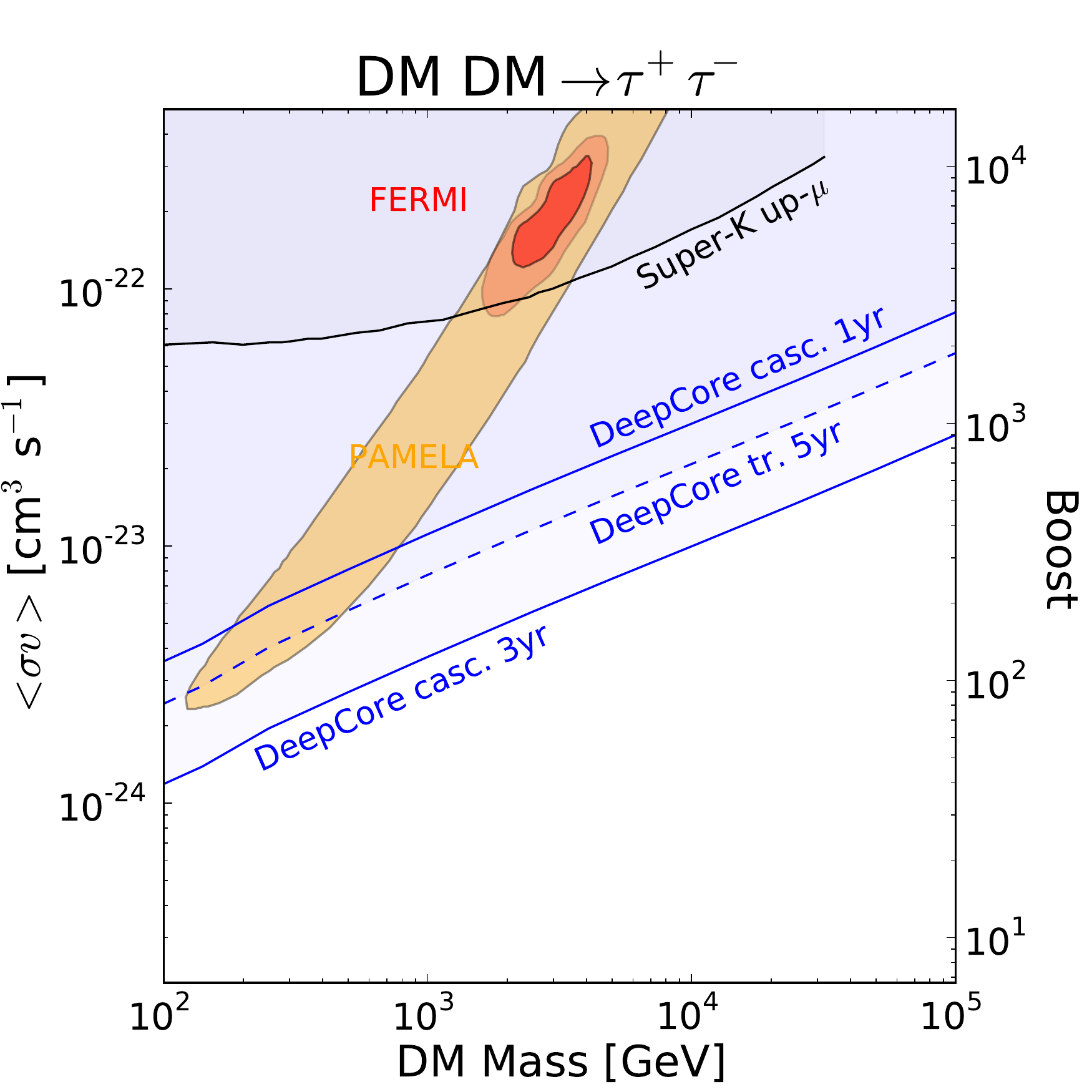}
\end{center}
\caption{Constraints for annihilation to $\mu^+\mu^-$ (left) and $\tau^+\tau^-$
(right); the regions above the contours are excluded.
The black contour (``Super-K up-$\mu$'') is the 
Super-Kamiokande limit to $3\sigma$ from
up-going muons, the orange band
is the PAMELA-preferred region, and the red ellipses 
are the Fermi-preferred region; these three are given 
by Ref.~\cite{Meade:fermi}.
The dashed blue line (``DeepCore tr. 5yr'') is the constraint to $2\sigma$ 
from IceCube+DeepCore for $\nu_\mu$ track-like 
events after five years of running, and the solid blue lines 
are the constraints to $2\sigma$
for all-flavor cascade events after one year (``DeepCore casc. 1yr'') 
and three years (``DeepCore casc. 3yr'') 
of running.}
\label{ann}
\end{figure}
The results for annihilation are shown in Fig.~\ref{ann}; the regions 
above the contours are excluded.  The plots 
show the same constraints as for decay, except that no isotropic limits are shown 
because they are weaker than the Super-Kamiokande limit by a factor
$\sim 10^5$ due to the low density of dark matter on cosmological scales.  
As with decays,
cascade events greatly accelerate the development of a useful constraint.
For the $\mu^+\mu^-$ final state the region by the Fermi data
can be eliminated in only
one year.

The exclusion plot for annihilations to $\tau^+\tau^-$ is shown only 
for completeness,
as the Fermi-preferred region has already been eliminated by the Super-Kamiokande
observation of upward-going muons.  However, it may provide a useful generic 
constraint on all dark matter models irrespective of the $e^\pm$ anomalies.

\section{Conclusion}
We have shown that, by using cascade events, 
IceCube+DeepCore can more quickly establish constraints
on dark matter models that would explain the reported $e^\pm$ anomalies, and
over time establish stronger constraints than from track-like events.
Specifically, track-like events will be able to significantly constrain the 
parameter space of decays
to $\mu^+\mu^-$, and rule out decays to $\tau^+\tau^-$ 
and annihilations to $\mu^+\mu^-$ in less than five years of running.  In comparison, 
cascade events can rule out decays to $\mu^+\mu^-$ in only three years,
 and rule out decays to $\tau^+\tau^-$ and annihilation to $\mu^+\mu^-$ 
 after only one year.  Moreover, these constraints 
 are highly 
 robust to the choice of dark matter halo profile and independent of
 dark matter-nucleon cross section.

In closing, we note two interesting possibilities for future work.  First, if the pointing accuracy
for track-like events at IceCube+DeepCore is established to be less than $10^\circ$,
the signal-to-noise for annihilations may be significantly enhanced by observing
a smaller region around the galactic center, possibly out-performing
 cascade searches (albeit with greater dependence on the
choice of dark matter halo profile).  This would strengthen the discovery
potential for dark matter because the galactic center could be identified
as a localized source of excess neutrinos.   Second, because 
 the $\nu_\tau$
atmospheric background is so low at energies
above 40 GeV and at low zenith angles,
if IceCube+DeepCore can demonstrate efficient
$\nu_\tau$ discrimination~\cite{grant:nutau}, 
signal to noise could be increased
by a factor $\sim 100$.  This would put leptophilic dark matter to a severe test.

\section{Acknowledgments}
The authors would like to thank D.~Grant, D.~J.~Koskinen, and I.~Taboada
for answering questions about DeepCore.

SKM is supported by World Premier International Research Center  
Initiative (WPI Initiative), MEXT, Japan, and would also like to thank
G.~Lambard for useful discussions. 
MRB is supported by the U.S. Department of Energy, under Grant No. DE-FG03-92-ER40701.
KF is supported by the U.S. Department of Energy and MCTP via the 
University of Michigan. DS is supported by
NSF Grant No. AST-0507117 and GAANN (D.S.).  HM is supported in part by World 
Premier International Research Center Initiative (WPI Initiative), MEXT, Japan,
in part by the U.S. DOE under Contract No. DE-AC03-76SF00098, and in part by 
the NSF under Grant No. PHY-04-57315.






\end{document}